\documentclass[12pt]{article}
\RequirePackage{amsthm,amsmath, amssymb}
\usepackage{graphicx,psfrag,epsf}
\usepackage{enumerate}
\usepackage{natbib}
\usepackage{algorithm}
\usepackage{algpseudocode}
\usepackage{soul}

\makeatletter
\renewcommand{\ALG@name}{Pseudo-algorithm}
\makeatother

\DeclareMathOperator*{\argmax}{arg\,max}
\newcommand{\blind}{0}

\begin{document}

\def\spacingset#1{\renewcommand{\baselinestretch}%
{#1}\small\normalsize} \spacingset{1}

%%%%%%%%%%%%%%%%%%%%%%%%%%%%%%%%%%%%%%%%%%%%%%%%%%%%%%%%%%%%%%%%%%%%%%%%%%%%%%

\if0\blind
{
  \title{\bf Nonparametric semisupervised classification with application to signal detection in high energy physics}
  \author{Alessandro Casa\hspace{.2cm}\\
    Department of Statistical Sciences, University of Padova\\
    and \\
    Giovanna Menardi \\
    Department of Statistical Sciences, University of Padova}
  \maketitle
} \fi

\bigskip
\begin{abstract}
Model-independent searches in particle physics aim at completing our knowledge of the universe by looking for new possible particles not predicted by the current theories. Such particles, referred to as \emph{signal}, are expected to behave as a deviation from the \emph{background}, representing the known physics. Information available on the background can be incorporated in the search, in order to identify potential anomalies. 
From a statistical perspective, the problem is recasted to a peculiar classification one where only partial information is accessible. Therefore a semisupervised approach shall be adopted, either by strengthening or by relaxing assumptions underlying clustering or classification methods respectively. 
In this work, following the first route, we semisupervise nonparametric clustering in order to identify a possible signal. 
The main contribution consists in tuning a nonparametric estimate of the density underlying the experimental data with the aid of the available information on the physical theory. As a side contribution, a variable selection procedure is presented. The whole procedure is tested on a dataset mimicking proton-proton collisions performed within a particle accelerator. 
While finding motivation in the field of particle physics, the approach is applicable to various science domains, where similar problems of anomaly detection arise.
\end{abstract}

\noindent%

\spacingset{1.45}
\section{Introduction}
\label{sec:intro}

\subsection{Framework and motivation}
Since the early Seventies, the \emph{Standard Model} has represented the state of the art in High Energy Physics. 
It describes how the fundamental particles interact with each others and with the forces between them (electromagnetic, weak and strong nuclear forces), giving rise to the matter in the universe. Within the Standard Model, a pivotal role is played by the Higgs boson, which imparts mass to some fundamental particles that would otherwise be massless. 
While its recent empirical confirmation \citep{Atlas_Higgs:12, CMS_Higgs:12} has represented an essential step to prove the consistency of the Standard Model, there are  indications that the current dominant theory does itself not complete our understanding of the universe. In fact, it fails to explain some phenomena as gravity, the nature of dark matter, as well as the dark energy, the last one accounting by itself for about the two thirds of the universe. 

All those attempts aiming to explain the shortcomings of the Standard Model go under the heading of \emph{Physics Beyond the Standard Model}. 
Experiments are conducted within large particle accelerators such as the LHC at Cern, where particles are made collide and the product of their collisions detected.
Research in this context is often performed in a \emph{model-dependent} fashion, trying to confirm some alternative physical conjectures (e.g. the so-called \emph{Supersymmetry}). In this work we follow, conversely, a \emph{model-independent} approach, not constrained to any specific physical theory already formulated. Model-independent searches aim at detecting empirically any possible \emph{signal} which behaves as a deviation from the \emph{background} process, representing, in turn, the known physics.  

From a statistical perspective, the considered problem is naturally recasted to a classification framework, although of a very peculiar nature. While the background process is known and a sample of virtually infinite size can be drawn from it, the signal process is unknown, possibly even missing. Available data have, consequently, two different sources: a first sample from the background process, generated via Monte Carlo simulations which mimic the results of collisions under the Standard Model; and a second sample of experimental data, drawn from an unknown generating mechanism, which surely include observations from the background but might also include observations from the signal. Due to the different degree of knowledge of the underlying generating processes, the two samples are referred to as \emph{labeled} and, respectively, \emph{unlabeled}. 
%a first, labelled, sample generated from the background process via Monte Carlo simulations which mimic the results of the collisions and account for the detector measurement uncertainty, and a second, unlabelled experimental sample, assumed to be drawn from an unknown generating mechanism, which surely include observations from the background but might also include observations from the signal. 

Hence, a semisupervised perspective shall be adopted, to gain knowledge from data for which only partial information is available \citep{semisupl}. %\citep{Kumar}?. 
In principle, and depending on the nature of the partial knowledge, semisupervised methods are built either by relaxing assumptions and requirements of supervised methods, or by strengthening unsupervised structures through the inclusion of the additional information available.  
We follow the latter route, in a nonparametric guise, as such formulation appears particularly consistent with some physical notion of signal. A common assumption in High Energy Physics is that a new particle would manifest itself as a significant peak emerging from the background process, in the distribution of the particle mass reconstructed from the available data. 
\emph{Nonparametric} (\emph{modal}) clustering, in turn, draws a correspondence between groups and the modal peaks of the density underlying the observed data, since clusters are defined as dense regions of the sample space. Thus, the one-to-one relationship between clusters and modes of the distribution provide an immediate physical meaning to the detected clusters. 

Further reasons make this approach to unsupervised learning appropriate in the considered context. Linking the groups to specific features of the probability distribution assumed to underlie the data allows to frame the clustering problem into a standard inferential context. Hence, it is possible both to estimate the number of clusters, and to resort to formal testing procedures. Both the tasks, typically prevented by alternative unsupervised approaches, are especially favorable in the physical context: groups may be labeled as background by exploiting the knowledge of the process or, by elimination, as signal, and any signal claim can be motivated by statistical evidence, as required by scientific discoveries. Furthermore, associating clusters to the characteristics of a probability distribution allows to partition the whole sample space, in addition to the observed data. This trait can be exploited to classify observations deriving from new experimental settings and not employed in the estimation phase, as it will be clarified in the next sections.  
Finally, the modal 
notion of cluster is not linked to any specific cluster shape, and employing a nonparametric approach to estimate the density allows for preserving this freedom operationally. Considering the physical framework where a possible signal is completely unknown, it would be indeed unrealistic to assume a predetermined shape for it.
%Further reasons make this approach to unsupervised learning appropriate in the considered context. First, it relies on a precise statistical notion of cluster, associated to a specific feature of the probability distribution assumed to underlie the data. This entails on one side that the number of clusters is conceptually well-defined and, then, operationally estimable. 
%A fair knowledge of the background process may allow for labelling groups as background itself or, by elimination, as signal. \hl{On the other hand, the definition of a probability density function underlying the data allows for the use of standard inferential tools to evaluate a clustering configuration, as it is required in a physical context where any claim shall be motivated by a strong evidence.}  
%Second, the modal
%notion of cluster is not linked to any specific cluster shape, and employing a nonparametric approach to estimate the density allows for preserving this freedom operationally. Considering the physical framework where a possible signal is completely unknown, it would be indeed unrealistic to assume a predetermined shape for it.
%Finally, since clustering is induced by the underlying density function, a partition of the whole sample space is possible. This trait can be exploited to classify observations deriving from new experimental settings and not employed in the estimation phase, as it will be clarified in the next sections.  

Within the described framework, this paper introduces a nonparametric global methodology aimed at looking for the possible presence of signals which exhibit as high-density peaks in the estimated distribution underlying a set of unlabeled data. The methodology is designed to integrate, within a nonparametric clustering formulation, the additional information we have about the background labeled process. Two main contributions can be highlighted: under the assumption that a signal does exist, the main idea is to tune a nonparametric estimate of the unlabeled data by selecting the smoothing amount so that the induced modal partition, where the signal emerges as a bump, classifies the labeled background data as accurately as possible.
%the main idea is to tune a nonparametric estimate of the unlabelled data, by selecting the amount of smoothing so that the induced modal partition classifies labelled background data as accurately as possible. 
%The induced signal, emerging by construction as a bump in the background distribution unseen in the distribution of the labelled data, either is spure  
Any significance of the forced bump would provide empirical evidence of a signal, and should then represent the stepping-stone to further investigate the physical properties and behaviour of the detected anomaly, for the possible claim of new physics discovery. As a second, side contribution, we propose a variable selection procedure, specifically conceived
%Since a signal is expected to emerge as a bump in the background distribution, its identification, unseen in the distribution of the labelled data, would provide \hl{a first empirical} evidence of a signal. \hl{It may thus represent a first step to further investigate the physical properties and behaviour of the detected anomaly, for the possible claim of new physics discovery}. As a second, side contribution, we propose a variable selection procedure, specifically conceived
for this framework, again exploiting available information about the background process. This procedure allows us to work in a lower dimensional space, where nonparametric methods provide more accurate estimates and where more interpretable results can be obtained.

The paper is organized as follows. After providing an overview of the literature inherent to the considered problem (Section \ref{sec:literature}), we outline 
the nonparametric approach to clustering (Section \ref{sec:modal_clustering}). Then, we propose the semisupervised nonparametric methodology for signal detection (Section \ref{sec:proposta}), and illustrate its application to a set of physical data (Section \ref{sec:analisi}). A discussion concludes the paper (Section \ref{sec:discussion}).   

\subsection{Related literature}\label{sec:literature}
%\subsection{Anomaly detection and semisupervised learning}

The peculiarity of the considered problem makes not trivial the inspection of the inherent literature. In fact, the aim of discriminating a possible signal which is expected to have an anomalous behavior with respect to the known background, frames into an \emph{anomaly detection} problem. 
Examples of such situations %are the analysis of electrocardiogram data, data from hacker attacks from
%remote machines, host-based intrusion detection systems, credit card data including fraudolent transactions. 
can be found in several domains. In networking, automatic systems are required to detect host-based intrusions. Similarly, in banking, credit institutes want to detect and prevent out of patterns of fraudulent spendings. In manufacturing, it is of interest detecting abnormal machine behaviors to prevent cost overruns, while in medical analysis detection of anomalies may be the sign of some disease. In all these situations, the normal behavior of the process of interest can be considered as known, since data are usually available in large amounts; conversely, tools to analyze unprocessed data which might include anomalies are required.  

In the considered setting, anomalies are expected to lie
within the domain of the background data, hence a single signal observation would look as if it was produced by the background process.
For this reason, anomalies are not to be searched among individual observations, but it is their occurrence together as a collection to be considered anomalous, and then possibly indicative of a new unknown particle. The problem is sometimes refereed to as \emph{collective anomaly detection} \citep[see, e.g.][]{Chandola_etal:09}.

For the analysis of such data, the presence of time, spatial, or some other kinds of relationships between observations is exploited to identify the anomalous regions; distance-based or, in general, clustering methods are typically employed otherwise. 
%is often particularly challenging as it involves exploring structure in the data for anomalous regions.

Staying within the field of High Energy Physics, anomaly detection has been often driven by the assessment of the degree of compatibility between two samples, and conducted via hypothesis testing \citep[e.g.][]{nps1, inverse_bagging}. Alternatively, \citet{nps2} have employed unsupervised, or weekly supervised neural networks to search for new physics, with some analogies with the approach followed in this work. 
A specific contribution that is worth to mention is the one of \citet{vatanen2012}, where a clustering-based semisupervised approach relying on parametric assumptions is proposed to face the same problem as the one considered here. 
%The main assumptions under which the method proposed works: (1) availability of two sample, $\mathcal{X}_{b}$ and $\mathcal{X}_{bs}$, respectively from \emph{background} and from the unknown, potentially containing the signal, process, (2) there is enough anomalies in order to enable inference, (3) the dimensionality of the data can be reduced to allow mixture modelling, (4) the \emph{background} has stationary distribution.
The authors propose a modification of the Expectation-Maximization algorithm \citep{EM} to estimate the probability density function underlying the experimental data, specified as a mixture of parametric distributions. One component of such mixture, describing the background density, is estimated based on the background data only; the other component, representing the possible signal, along with the mixing proportions, are estimated in a second step based on the experimental data. A goodness of fit test serves to discard insignificant components and assures that the whole estimated density is equal to the background component when no signal is detected. 
   
%The authors propose a modification of the Expectation Maximization algorithm \citep{EM} to estimate the probability density function $f_{bs}$, which is specified as follows: %so-called \emph{fixed background model} defined as 
%\begin{equation}
%\begin{split}
%$$f_{bs}(x)=\pi_{b}f_{b}(x)+\pi_{s} f_{s}(x) $$%\\
%&=\pi_{b}f_{b}(x)+\sum_{q=J+1}^{J+Q} \pi_{q}N(x|\mu_{q},\Sigma_{q}).\\
%\end{split}
%\end{equation}

%The background density is es
%The \emph{background} density $f_b$ is specified as a Gaussian mixture of $J$ components, and estimated via maximum likelihood based on the labelled data $\mathcal{X}_b.$ 
%Afterward, the mean and the covariance parameters of $\hat f_{b}$ are kept fixed, while the weights and parameters of the new possible component $f_{s}$, indicating a possible signal, are iteratively estimated on the basis of the unlabelled data $\mathcal{X}_{bs}$. A goodness of fit test on $f_{bs}$ then serves to discard not significant components of $f_{s}$ and assures that $f_{bs}$ will be equal to $f_{b}$ in the case where no observations from the \emph{signal} process are detected. 

\section{Nonparametric clustering}
\label{sec:modal_clustering}
\emph{Nonparametric} or \emph{modal} clustering delineates a class of methods for grouping data defined on a topological, continuous space, and built
on the concept of clusters as ``continuous, relatively densely populated regions of the space, surrounded by continuous, relatively empty regions"
\citep{Carmichael_etal68}.
 
\begin{figure}[bt]
\includegraphics[width=0.33\textwidth, height=4.5cm]{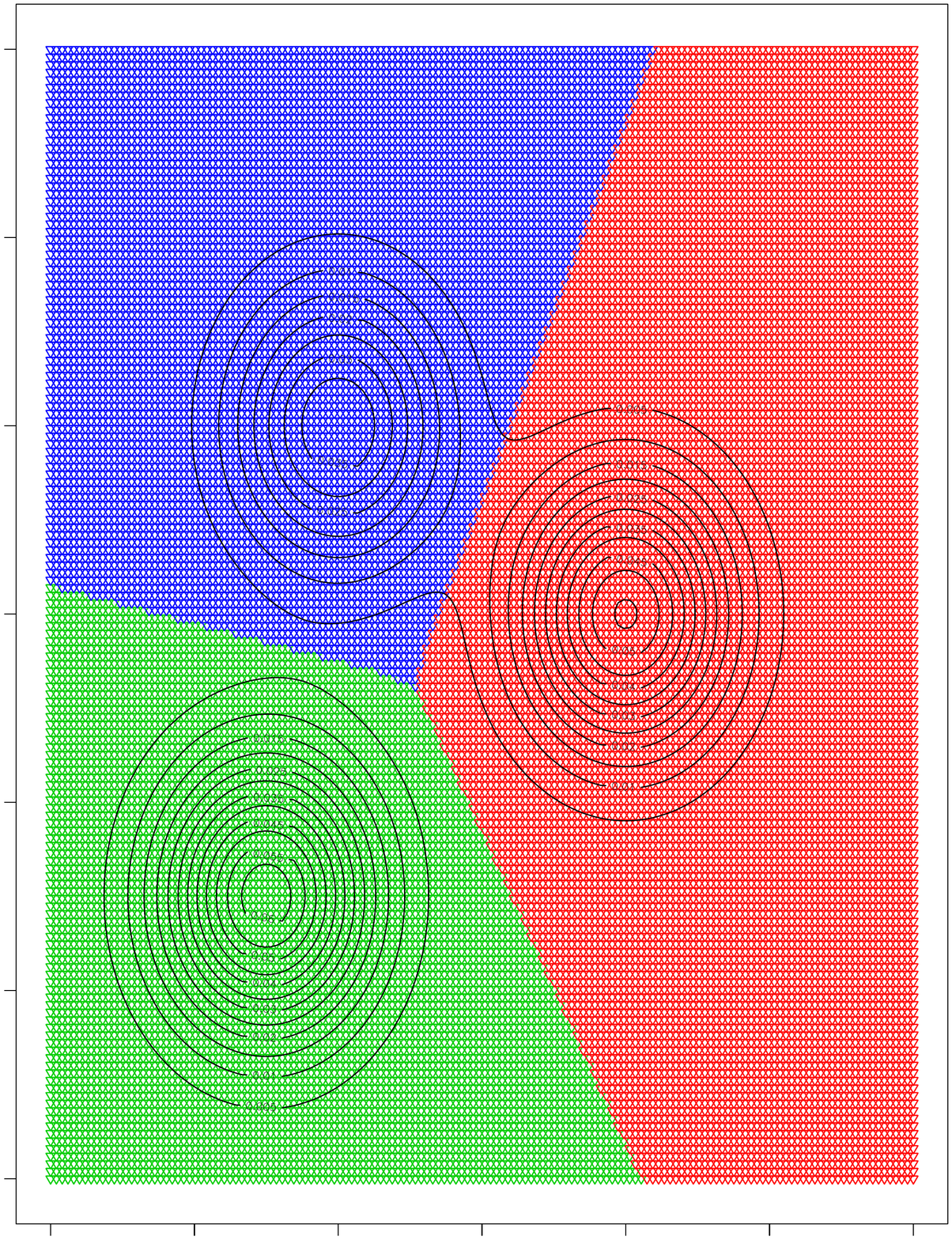}%\hspace{0.03cm}
\includegraphics[width=0.33\textwidth, height=4.5cm]{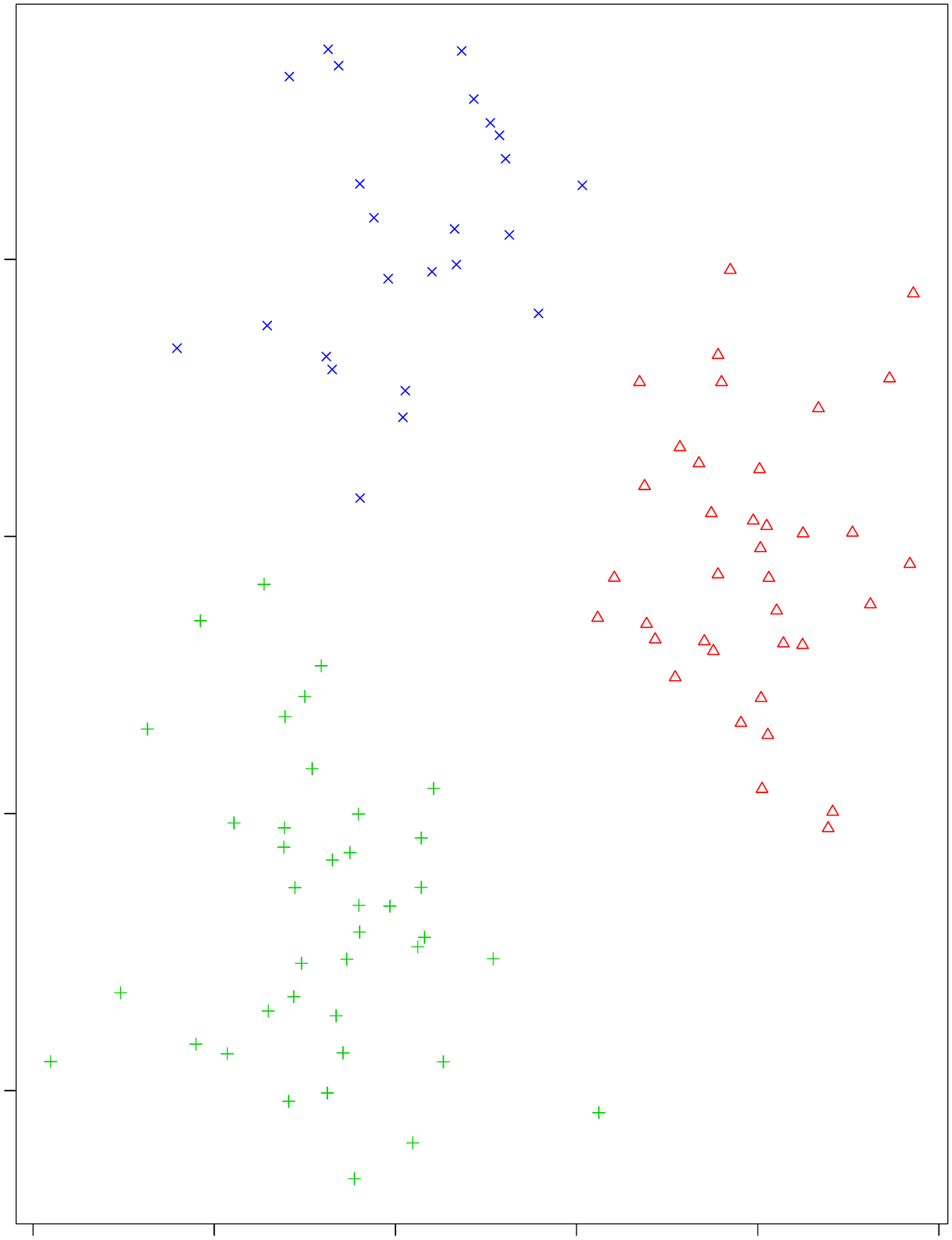}
\includegraphics[width=0.33\textwidth, height=4.5cm]{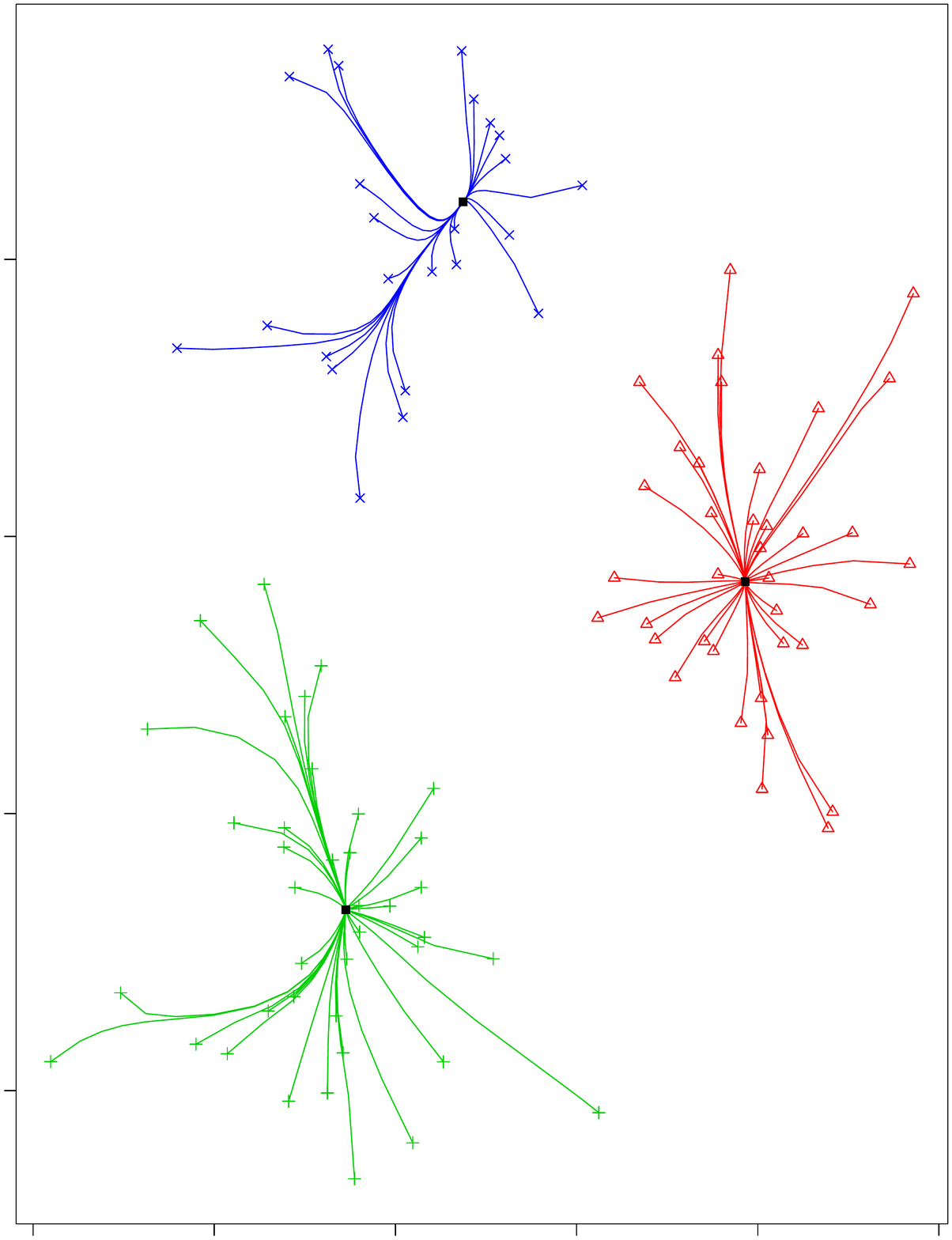}
\caption{A trimodal density and the sample space partition induced by domain of attraction of the modes; 
a sample generated from the density, and the path of each observation to climb the gradient towards the estimated modes according to the mean-shift clustering.}\label{fig:modal}
\end{figure}

The observed data $\mathcal{X} = \{\mathbf{x}_{i}\}_{i=1, \ldots, n},$ $\mathbf{x}_{i}\in \mathbb{R}^d$ are supposed to be 
a sample from a multidimensional random variable with (unknown) probability density function $f$. 
The modes of $f$ are regarded to as the archetypes of the clusters, which are in turn represented
by their domains of attraction.
This idea has found a proper formalization in \citet{chacon2015}. By
exploiting some notions from differential topology, the author defines a cluster as the
unstable manifold of the negative gradient flow corresponding to the local maxima of $f$.
Intuitively, if $f$ is figured as a mountainous landscape, and modes are its peaks, clusters are
the “regions that would be flooded by a fountain emanating from a peak of the mountain
range”. These notions are illustrated for a bivariate example in Figure \ref{fig:modal}.
Note that since the groups are induced by the gradient of the underlying density, clustering is not limited to the observed points, but can be extended to any point of the sample space. 

Operationally, clustering involves two main choices, which are overviewed in the following. See \citet[]{menardi2016} and references therein for further details.  

The first choice concerns the operational identification of the modal regions,
which may occur according to different paradigms. 
One strand of methods, searching directly for the modes of $f$, naturally comply with the previously outlined notion of cluster. Most of the contributions following this direction can be considered as refinements of the mean-shift \citep{Fukunaga_Hostetler75}, an iterative mode-seeking algorithm that, at each
step, moves the data points along the steepest ascent path of the gradient, until converging to a mode. Operationally a partition of the data points is obtained by grouping in the same cluster those observations ascending to the same mode of the density; the right panel of Figure \ref{fig:modal} provides a simple illustration of this idea.
A second strand of methods does not attempt the task of mode detection
but associates the clusters to disconnected density level sets of the sample space, as the
modes correspond to the innermost points of these sets. 

The second choice concerns the estimation of the density function, which determines the high density regions and, hence governs the final clustering. 
Which specific estimator is employed depends on either conceptual or operational convenience reasons, but the selection usually falls  
within a nonparametric formulation. Disregarding the specific choice adopted, nonparametric
 estimators depend on some parameters defining the amount of smoothing. 
Consider, for example, a product kernel estimator, specified as follows 
\begin{equation}\label{eq:kernel}
\hat{f}(\mathbf{x}; \mathcal{X}, h) = \frac{1}{n\cdot h^d} \sum_{i=1}^n \prod_{j=1}^d K\left(\frac{x_{ j}-x_{ij}}{h}\right),
\end{equation}

where $K$ is a symmetric probability density function and $h > 0$ is the bandwidth which defines the degree of smoothing. 
How to set this parameter is an issue to be tailored, as it affects both the shape and the number of modes 
of the estimate: a large $h$ oversmooths the density function thus averaging away features in the highest density regions,
 while a small $h$ undersmooths the density by favoring the appearance of spurious modes.   
The number of nearest neighbors plays a similar role in $k-$nearest neighbor
estimates, and the number of summands determines the amount of smoothing 
in orthogonal series estimators. 

\section{Nonparametric semisupervised learning}\label{sec:proposta}

\subsection{Notation and assumptions}
In the rest of the paper, we adopt the following notation: $\mathcal{X}_{b} =\{\mathbf{x}_i\}_{i=1,\ldots,{n_b}}$ 
denotes the set of labeled data, supposed to be a sample of i.i.d. realizations from the background distribution $f_b.$ 
Each observation represents a collision event recordered by a detector within a proton-proton collider. 
As such, $\mathbf{x}_i =(x_{i1}, \ldots, x_{ij}, \ldots, x_{id})'$ corresponds to different characteristics of the topology of a collision event (e.g. the number of tracks, the high-transverse momenta of new objects produced by the collision, etc);   
$\mathcal{X}_{bs} =\{\mathbf{x}_i\}_{i=1,\ldots,n_{bs}}$ has the same structure and denotes, in turn, the unlabeled set of data, assumed to be drawn from the whole underlying distribution $f_{bs}. $ %As far as there exists new undiscovered physics beyond the Standard Model, $f_{bs} $ and $f_b$ are different just because of the presence of a signal which features as a new mode of $f_{bs}, $ not detected in $f_b.$

Similarly to \citet{vatanen2012}, our work relies on the following assumptions: (i) as far as a signal exists, it arises as a new mode in $f_{bs}$, not seen in the background density $f_b$;
(ii) it arises in a fraction of data that is large enough to enable collective inferences; (iii) its underlying structure is revealed in a lower-dimensional space with respect to the one defined by all the observed variables; 
(iv) the background has a stationary distribution, i.e. the Monte Carlo sample $\mathcal{X}_b,$ perfectly captures the true distribution of the background, and $f_{b}$ possibly differs from $f_{bs}$ just because of the presence of a signal.

A few comments shall be pointed out to discuss the assumptions above. First, it may be the case that a physical signal would not exhibit as a new mode unseen in the background density. For example, it could lie on top of an existing mode and simply raise that mode. However, the assumed behaviour is common for several signals \citep[see, e.g.][]{librofisica} and this somewhat strong assumption allows us to gain more power with respect to the searches for undisclosed kinds of anomalies. Second, since new undiscovered physics would be certainly rare, it might appear unrealistic to assume the observation of a sensible fraction of signal events. In fact, in physical applications whole regions of the sample space are completely drop out from the analysis as known not to bear useful information; focusing specifically on a subset of the domain implies an automatic increase of the frequence of observations coming from a possible signal. As far as assumption (iii) concerns, it is largely reported that a typical aspect of high-dimensional data is the tendency to fall into manifolds of lower dimension \citep{scott}. Finally, while Monte Carlo simulations of the background often require simplifications and approximations and can be a major source of errors, it is quite common to assume that the main behaviour of the process is, in fact, catched, and we may rely on that to investigate whether $f_{bs}$ is, in fact, different from $f_{b}.$ 

\subsection{On choosing the amount of smoothing}
Due to the key role played by the density in nonparametric clustering, it makes sense to semisupervise the learning process by strenghtening it via the inclusion of the additional information available on the labeled data within the phase of density estimation. 

Whatever specific estimator is selected, one of the most critical aspects of nonparametric density estimation consists in tuning some parameter which governs the amount of smoothing and, hence, the modal structure. In the following, we focus on the product kernel estimator (\ref{eq:kernel}), but the methodology easily applies to other choices of estimators. 

The idea underlying the proposed procedure is to %exploit the available information on the background process by 
identify the modal partition of the experimental, unlabeled, data induced by the density estimate which guarantees a signal warning while allowing for an accurate classification of the background labeled data. %Albeit reasonable in principle, in the considered application this approach requires some slight modifications since otherwise it would mask completely any possible signal, hence preventing the usefulness of the proposal, by suggesting the same modalities for $f_b$ and $f_{bs}$.  

%For this reason, and coherently with previous considerations about the pre-processing and filtering nature of the procedure, we take a slightly different route. We aim indeed at highlighting changes in the modal structure of $f_b$ and $f_{bs}$, as they would be flags of a possible signal, simultaneously classifying background observations as good as possible conditionally to those changes. 

Specifically, let $\hat{f}_{b}$ be an estimate of $f_{b}$, which we may consider to be arbitrarily accurate due to the availability of as many data as required from the background process (see below for a discussion about this aspect). The estimate $\hat{f}_{b}$ induces a partition $\mathcal{P}_{b}(\mathcal{X}_{b})$ of the background data $\mathcal{X}_b$, determined by its modal structure. Then, for a grid of bandwidths $h_{bs}$ varying in a range of plausibile values, the estimates $\hat{f}_{bs}(\cdot;\mathcal{X}_{bs},h_{bs})$ of the whole process density $f_{bs}$ are obtained, each of them inducing a partition $\mathcal{P}_{bs}(\mathcal{X}_{bs})$ of the unlabeled data $\mathcal{X}_{bs}$, as well as a partition of the sample space, defined by its modal regions. The latter partition allows to determine the cluster membership of $\mathcal{X}_b$, i.e. a partition $\mathcal{P}_{bs}(\mathcal{X}_{b})$. 
The two partitions $\mathcal{P}_{b}(\mathcal{X}_{b})$ and $\mathcal{P}_{bs}(\mathcal{X}_{b})$, induced respectively by the modal structure of $\hat{f}_{b}(\cdot;\mathcal{X}_{b},h_{b})$ and $\hat{f}_{bs}(\cdot;\mathcal{X}_{bs},h_{bs})$, can then be compared via the computation of some agreement index $I$. 

Assume, without loss of generality, that high values of $I$ indicate an agreement between $\mathcal{P}_{b}(\mathcal{X}_{b})$ and $\mathcal{P}_{bs}(\mathcal{X}_{b})$. 
Then, the ultimate partition $\mathcal{\tilde P}_{bs}(\mathcal{X}_{bs})$ of the unlabeled data will be induced by the estimated density $\hat{f}_{bs}(\cdot;\mathcal{X}_{bs},\tilde h_{bs})$, built on the \emph{best undersmoothing bandwidth}, i.e.
\begin{equation}\label{eq:hbs}
\tilde h_{bs} = \argmax_{h_{bs} \in \mathcal{H}} I(\mathcal{P}_{b}(\mathcal{X}_{b}),\mathcal{P}_{bs}(\mathcal{X}_{b})) \; , 
\end{equation}
where $\mathcal{H}=\{ h_{bs} : \mathcal{M}_{bs} > \mathcal{M}_b\}$ and $\mathcal{M}_{bs}, \mathcal{M}_b$ represent the number of modes of $\hat{f}_{bs}(\cdot; \mathcal{X}_{bs},h_{bs})$ and $\hat{f}_{b}(\cdot; \mathcal{X}_{b},h_{b})$ respectively. The significance of the $\mathcal{M}_{bs} - \mathcal{M}_b$ additional modes then becomes the focus to investigate on, to eventually decide whether proceed with further investigations about a possible signal claim and its features. %This partition is then used as a starting point to formally test the relevance of the possibly detected signal and to further investigate on its features.

To better figure out the evolution of the agreement index as a function of $h_{bs}$, it is possible to highlight some recurring behaviours. Small values of $h_{bs}$ will determine an indented $\hat{f}_{bs}$, not compatible with the clusters of $\mathcal{X}_b$. For increasing $h_{bs},$ $\mathcal{M}_{bs}$ is expected to decrease, eventually leading to a unimodal structure, as the number of modes of a density estimate is grossly decreasing with the amount of smoothing. This is precisely true for kernel estimators with some specific choices of kernels. %On the other hand, a large $h_{bs}$ will tend to oversmooth $f_{bs},$ eventually leading to a unimodal structure. 
The associated behaviour of $I$ will depend on the characteristics of $f_b$. If, with a multimodal background, large values of $h_{bs}$ will not reflect the clustering structure of $f_b$ and lead to a decreased $I$, in the unimodal case the agreement index will grows with $h_{bs}$ and a perfect recovering of the background partition will be achieved just as an effect of oversmoothing $f_{bs}$, disregarding the presence of a signal. %Given the assumption about the signal emerging as a new mode in the background density, both the cases lead to a dead end when signal detection is the final aim of the analysis.  

The ground underlying the proposed procedure is that the background process is dominant with respect to any possible signal, and its density estimate $\hat f_b$ is arbitrarily accurate. Since $f_b$ and $f_{bs}$ are assumed to differ just because of the possible presence of a signal, it appears sensible to preserve a good characterization of the background features by inducing an agreement between $\mathcal{P}_{b}(\mathcal{X}_{b})$ and $\mathcal{P}_{bs}(\mathcal{X}_{b})$. On the other hand, yet for the prevalence of the background process, its features are going to be largely persistent across different smoothing amounts. %Even in the presence of a signal, a perfect agreement of the two partitions would be anyway achieved for some $h_{bs}$ just as an effect of oversmoothing the density, by ``cleaning out'' the evidence of a signal. 
In fact, by choosing the amount of smoothness implied by the best undersmoothing bandwidth, %where the optimality is driven by the agreement index, 
we aim at preserving as much as possible the relevant structures of the background process, while highlighting new modes. Whether these modes are actually candidate to be a signal or just sampling artifacts is then established by formally testing their significance and further investigating on their features.

%The proposed approach follows, and simultaneously adapts, this rationale to the nature of the procedure that is conceived as a pre-processing tool to filter events in order to highlight observations more likely to be labelled as signal. By choosing the amount of smoothness implied by the \emph{best undersmoothing bandwidth}, where the optimality is driven by the agreement index, we aim at preserving as much as possible the relevant structures of the background process while highlighting new modes whose behaviour has to be studied further. 

For an operational description of the procedure see the Pseudo-algorithm \ref{alg:selhbs}.

\subsubsection{Remarks} 
The actual implementation of these ideas requires a number of operational choices, discussed in the following. 
\begin{itemize}
\item As a first step the procedure requires an estimate of the background density $f_{b}$, and therefore the selection of an appropriate smoothing parameter $h_{b}$. In the specific application, the complete knowledge of the background process, and the consequent availability of an arbitrarily large number of observations drawn from it, makes the selection not critical. %allowing to choose among a plethora of selectors in order to obtain a working bandiwdth. 
%has been shown \citep[see, e.g.][]{wand1994kernel} that, %with increasing sample size and 
Under minimal regularity assumptions, the kernel density estimator is consistent, hence with a huge sample size small changes in the quality of the estimate are expected by selecting $h_b$ via any sensible automatic selector proposed in literature;
%For what we have said, it is suggested to obtain $\hat{f}_b$ by simply tuning the level of smoothness considering one of the bandwidth selctors proposed in literature;
see \cite{wand1994kernel} and references therein for further details.

\item The agreement index $I$ employed to select $h_{bs}$ may be any external validation index employed to compare different partitions of the same data. Sensible choices are, for instance, the \emph{Fowlkes-Mallows} coefficient, the \emph{Jaccard} index, the \emph{Adjusted Rand Index} \citep[][Ch. 27]{hennig_etal2015}.
%\item The comparison between $\mathcal{P}_{bs}(\mathcal{X}_{b})$ and $\mathcal{P}_{bs}(\mathcal{X}_{b})$ required to compute $I$ and select $h_{bs}$ is not performed on the whole background sample $\mathcal{X}_{b}$ but on a number of different subsamples $\mathcal{X}_{b}^{*}$ drawn from it. This allows to obtain the empirical distributions of the agreement index $I$ for varying $h_{bs}$, and the values of the median $I$ are then optimized with respect to $h_{bs}$ in order to robustify the choice.

\item Selecting the best undersmoothing bandwidth $\tilde h_{bs}$ entails the whole process density $f_{bs}$ to be estimated under the assumption of the presence of a signal. In fact, most of the physical experiments are expected to produce no signal, hence further steps are needed to establish whether the additional modes of $f_{bs}$ are spurious or actual candidate to be signals.

To this aim, various tools can be employed. According to the concept of persistent homology \citep{Fasy_etal2014}, non-spurious modes are generally associated with enduring behaviours for varying bandwidth. Hence, a true signal is expected to produce a plateau in the plot of agreement index versus the bandwidth. This idea is also related to the rational underlying the SiZer map \citep{sizer}, a graphical device 
to display significant features in univariate curves. Alternative tools to test mode significance, also working in the multidimensional context, have been proposed, among others, by \citet{peronepacifico, burman_polonik09, duong_etal08}. 

\end{itemize}

%Lastly note that the choices of the clustering method and of the agreement index does not represent a constraints and other road could be taken. 

%In the procedure $\mathcal{P}_{bs}(\mathcal{S}_{b})$ is obtained assigning each observation in $\mathcal{S}_{b}$ at the cluster for which it has higher estimated density; for this reason the estimation of the density of the clusters in $\mathcal{P}_{bs}(\mathcal{S}_{bs})$ is required. Although the nonparametric estimation of the densities of single clusters is straightforward, this is still an open problem in literature. It is known indeed that if the density of a cluster is estimated using only observations belonging to that cluster, we have an underestimation of the variance \cite{r7}. In this work, considering that the underestimation usually does not have a strong effect on clustering, we use for the estimation only the observations that belong to the cluster; nevertheless other approach and modification could be considered. 

\begin{algorithm}[t]
\caption{\emph{Semisupervised procedure for bandwidth selection}\protect\newline 
Denote with: %$\mathcal{X}_{b}$ the \emph{background} sample, $\mathcal{X}_{bs}$ the \emph{unlabelled} sample from the whole process.   
%Let $h_{b}$ be the \emph{background} bandwidth; $h_{bs}$ the whole process bandwidth (to be determined); 
$h_{grid}$ a grid of plausible values for $h_{bs}$; %$\mathcal{H}_{grid}$: the subset of $h_{grid}$ corresponding to bandwidths leading to a greater number of modes in the estimate $\hat{f}_{bs}$ with respect to $\hat{f}_b$. Finally let 
%$\mathcal{P}_{k}(\mathcal{X})$ be a partition of data $\mathcal{X}$ identified by the modal structure of density $f_{k}$ and 
$I(\mathcal{A}, \mathcal{B})$ an agreement index between partitions $\mathcal{A}$ and $\mathcal{B};$ $\alpha$ the $I-type$ error probability of testing the significance of the signal modes.  
}\label{alg:selhbs}
\begin{algorithmic}[1]
\Statex \textbf{Input} %\begin{itemize} \item 
$\mathcal{X}_{b},$ $\mathcal{X}_{bs},$ %\item 
$h_{b}$, $h_{grid}$, $\alpha$.
%\end{itemize}
\State compute $\hat{f_{b}}(\cdot; \mathcal{X}_{b},h_{b})$ and count its modes $\mathcal{M}_{b}$;
\State obtain $\mathcal{P}_{b}(\mathcal{X}_{b})$;
\State $\mathcal{H}_{grid} \gets \emptyset$
\For{h in $h_{grid}$}  
\State compute $\hat{f}_{bs}(\cdot; \mathcal{X}_{bs},h)$ and count its modes $\mathcal{M}_{bs}$
\If{$\mathcal{M}_{bs} > \mathcal{M}_b$}   
\State $\mathcal{H}_{grid} \gets \{\mathcal{H}_{grid}\cup h\}$ \EndIf  
\State obtain $\mathcal{P}_{bs}(\mathcal{X}_{b})$; 
\State compute $I\left(\mathcal{P}_{b}(\mathcal{X}_{b}), \mathcal{P}_{bs}(\mathcal{X}_{b})\right)$.
\EndFor
%\State $\tilde h_{bs} = \argmax_{h_{bs} \in \mathcal{H}} I(\mathcal{P}_{b}(\mathcal{X}_{b}),\mathcal{P}_{bs}(\mathcal{X}_{b}))$  
\State $\tilde h_{bs} \gets \argmax_{h\in \mathcal{H}_{grid}}  I\left(\mathcal{P}_{b}(\mathcal{X}_{b}), \mathcal{P}_{bs}(\mathcal{X}_{b})\right)$
\State compute $\hat{f}_{bs}(\cdot; \mathcal{X}_{bs},\tilde h_{bs})$;
\State test the significance $p$ of the modes of $\hat{f}_{bs}$
\If{$p < \alpha$} 
\State obtain $\mathcal{P}_{bs}(\mathcal{X}_{bs})$; 
\EndIf 
\vspace{0.3cm}
\Statex \textbf{Output}: $p;$ ($\mathcal{P}_{bs}(\mathcal{X}_{bs})$, if computed)
\end{algorithmic}
\end{algorithm}

\subsection{Variable selection procedure}
\label{sec:sec3.1}

Within a nonparametric framework, the curse of dimensionality is known to have a strong impact on the quality of the estimates. In the context of density estimation, for high dimensional sample spaces, much of the probability mass flows to the tails of the data density, possibly giving rise to the birth of spurious clusters and averaging away features in the highest density regions. 
Resorting to dimension reduction methods is then often advisable to work on a reduced subspace and improve the accuracy of the estimates.
The identification of the reduced subspace, either obtained by variable selection or by producing suitable combinations of the observed variables, is driven by the aim of preserving the relevance and the informativeness of the originally observed variables. However, defining the concepts of relevance and informativeness is not trivial in an unsupervised context, where the variables have a symmetric role because of the lack of a response one. 

In order to guarantee interpretation of the results in terms of the original measured variables, we pursue a variable selection approach. Subject-matter considerations may aid to define the concept of relevance and informativeness, which are then related to the aim of identifying a possible signal whose behaviour departs from the one of the background process. Additionally, dimensionality reduction may be driven (i.e. semisupervised) by taking advantage of the additional knowledge available on the background process. 

In this perspective, we assume a variable to be relevant if its distribution shows a changed behavior in $f_{bs}$ with respect to $f_{b},$ as this difference shall be only due to the presence of a signal, not seen in background density. 
This idea is pursued by comparing repeatedly the estimated densities $\hat{f}_{b}$ and  $\hat{f}_{bs}$ on subsets of variables and by eventually selecting those variables that, more often, are responsible for a different behavior of the two marginal distributions.

Among the many possible alternatives, we consider as a comparison criterion a two sample version of the \emph{integrated squared error}, extensively used in the nonparametric framework to assess the quality of a density estimate. The statistic, proposed by \citet{Anderson_etal1994}  to test the equality of two distributions and shown to be asymptotically normal \citep{Duong_etal2012}, is the \emph{integrated squared difference} between a kernel estimate of the two densities under evaluation. 
In our setting, the test is repeatedly applied on the marginal density kernel estimates of subsets of variables, based on the background and the whole process data. Formally, at each step $k$ variables are selected at random among the $d$ observed ones, the samples $\mathcal{X}_b$ and $\mathcal{X}_{bs}$ reduced coherently to $\mathcal{X}_b^k$ and $\mathcal{X}_{bs}^k$, and used to estimate the underlying distribution. Then the statistic  
\begin{equation}\label{eq:test}
\int_{\mathbb{R}^{d}}[\hat{f}_{b}(\cdot;\mathcal{X}_{b}^{k},h) - \hat{f}_{bs}(\cdot;\mathcal{X}_{bs}^{k},h)]^{2} dx
\end{equation}
is computed. Large values are considered evidence of a departure of $f_{bs}$ from $f_b,$ ascribable to a different behaviour of the selected $k$ variables. For those variables a counter is then updated to account for such evidence. At the end of the procedure, the counter will give an indication about the relative relevance of each single variable. If $d'<d$ variables show evidence of a remarkable relevance with respect to the other ones, these are selected and the associated reduced samples $\mathcal{S}_{b}$ and $\mathcal{S}_{bs},$ of size $n_{b}\times d'$ and $n_{bs} \times d'$ respectively, are then intended to be used in place of $\mathcal{X}_{b}$ and $\mathcal{X}_{bs}$ within the main methodology illustrated in the previous section.
%This procedure is intended to be applied as a preprocessing step of the methodology illustrated in the previous section. \\
%Hence in the following we will denote as $\mathcal{S}_{b}$ and $\mathcal{S}_{bs}$ respectively the background and the unlabelled sample having dimensions respectively $n_{b}\times d'$ and $n_{bs} \times d'$, where $d'<d$ is the number of variables selected by the procedure.

For an operational description of the procedure see Pseudo-algorithm \ref{alg:varsel}.

\subsubsection{Remarks} The procedure described so far, albeit in principle sensible, requires a few choices to be discussed. 
\begin{itemize}
\item In order to perform the test based on (\ref{eq:test}) under the null hypothesis of equal distributions, the kernel estimates of both the processes are built on the basis of the same bandwidth $h$. While, one more time, one has to deal with the problem of selecting such bandwidth, at this phase of the procedure any sensible bandwidth selector can be employed, as the main aim is not to obtain an accurate density estimate but just a fair comparison between the two distributions. % In the empirical analysis which will follow, we adopted the rule of thumb of selecting $h$ as asymptotically optimal for a Normal underlying background density.
\item While selecting at each step $k=1$ variables would guarantee to count for the relevance of the only variables possibly responsible for a different behaviour in the two processes under consideration, the choice of working with a subset of $k>1$ variables is due to the will of keeping relations among variables while working on a reduced space. It might occur, indeed, that a signal not emerging in the univariate behaviour of the observed variables would, in fact, manifest in their joint distribution. The choice of $k$ is subjective and it can be motivated both by theoretical considerations on the degradation of the estimates and by computational reasons. % In the empirical analysis of the next Section we selected $k=3$ variables. 
\item Operationally, a subset of $k$ variables is candidate to be relevant if the test based on (\ref{eq:test}) results in a low $p-$value. A possibile argument is that, using a test at each step of the procedure, a multiple testing problem arises. In fact, the procedure does not aim at testing the difference between distributions, but it is built just to give a general indication of which variables are responsible for a possible detected difference. In this sense, following heuristic, non-rigorous principles to a certain extent looks a sensible choice.
\item The proposed procedure to select the relevant variables implicitly assumes that the unlabeled data exhibit, in fact, the presence of a signal. When this is not the case, the relevance counter is likely not to vary that much across the variables, thus not showing evidence of some variables being more informative than others. This gives a first, rough, answer to the research question on whether the signal is present or not. 
\end{itemize}

%In literature it is possible to find different approaches to the this problem. Here we choose to reduce the dimension of the space selecting the variables considered to be the more relevant ones. The motivation behind this choice regards the application problem: we believe that, searching for a signal, it would be meaningful to be able to interpret the results in terms of the original measured variables and not in terms of suitable transformations.

\begin{algorithm}[t]
\caption{\emph{Semisupervised variable selection procedure}\protect\newline
Denote with: $M$ the number of iterations of the procedure and $k$ the number of variables selected at each iteration; %Let $\mathcal{X}_{b}$ and $\mathcal{X}_{bs}$ be the background sample and the unlabelled sample from the whole process, respectively having dimensions $n_{b} \times d$ and $n_{bs}\times d$. With $\mathcal{X}_{b}^{k}$ and $\mathcal{X}_{bs}^{k}$ we denote the corresponding samples on the $k$-dimensional selected subset. Let $\hat{f}_{b}(\mathcal{X})$ and $\hat{f}_{bs}(\mathcal{X})$ be the background and whole process densities estimated using the data $\mathcal{X}$. Finally 
\emph{count} a $d$-dimensional vector giving an indication about the relevance of each variable and $count_{k}$ the elements of \emph{count} indexed by the $k$ variables selected at that iteration. 
}\label{alg:varsel}
\begin{algorithmic}[1]
\Statex \textbf{Input} $\mathcal{X}_{b}$, $\mathcal{X}_{bs}$, $M$, $k$.
\State $count \gets (0, \ldots, 0)$
\For{i=1,\dots,\textit{M}}  
\State select randomly $k$ variables;
\State compare $\hat{f}_{b}(\mathcal{X}_{b}^{k})$ and $\hat{f}_{bs}(\mathcal{X}_{bs}^{k})$;
\If{$\hat{f}_{b}(\mathcal{X}_{b}^{k}) \ne \hat{f}_{bs}(\mathcal{X}_{bs}^{k})$}
\State update $count_{k} \gets count_{k}+1$.
\EndIf
\EndFor
\Statex \textbf{Output}: $k^{*}$, a vector of length $d'$ with $d'<d$, indexing the set of variables considered to be relevant. 
\end{algorithmic}
\end{algorithm}

\section{Application}\label{sec:analisi}
To illustrate the proposed methodology, we consider its application to a Monte Carlo physical process simulated at a parton level according to the configuration of the ATLAS detector, within the LHC at CERN. Note that unlike statistical simulations, where data are generated from a given probability distribution, physical Monte
Carlo simulations are realizations of possible collisions, produced by a complex system of subsequent steps, based on theoretical parton distribution functions, possible collision effects as, for instance,  creation of elementary particles, coupling, interactions. The simulations also cover further processes that the particles are subject to immediately after the collision, as for example their deceleration, scatterings or jet creation. For such simulated events, a detector response is computed based on its measurement efficiency, and a trigger is applied, to reflect conditions in
which the experimental data are collected. 

Each observation then corresponds to a single collision event and the associated variables describe the kinematics of the decaying results of the collisions. Variables are classified into 22 low-level features, representing basic measurements made by the particle detector, as well as the result of standard algorithms for reconstructing the nature of the collision, namely the leading lepton momenta, the missing transverse momentum magnitude and angle, the momenta of the first four more energetic jets, the b-tagging information for each jet. Additionally, 5 high-level variables are considered, which combine the low-level information to approximate the invariant masses of the intermediate particles. Since a few of the considered variables are highly discretized, they have been removed from the analysis to allow for a proper application of kernel methods. The final data count $d=23$ variables. 

The signal is simulated as a new particle of unknown mass which decays to a top quark pair production $t\overline{t}$. The known background is in turn a Standard Model top pair production, identical in its final state to the signal but distinct in the kinematic characteristics because of the lack of an intermediate resonance. Refer to \citet{data} for a detailed description of the data and their characteristics.

From the original data set including several millions of collision events of both the background and the signal processes, two samples $\mathcal{X}_{b}$ and $\mathcal{X}_{bs}$ have been drawn, each including $20000$ observations. In fact, the latter set has been splitted into two halves, to be used as training and test sample, so that $n_b = 20000$, $n_{bs} =10000$ ultimately, and a further test set $\mathcal{X}_{bs}^{T}$ of size $n_{bs}^{T}= 10000$ is at hand. The choice of the sample sizes is motivated by the aim of keeping the analysis computationally feasible with standard machines, but of course larger samples could be extracted, given the huge amount of data available, especially for the background. In the $\mathcal{X}_{bs}$ data set, we consider a signal proportion amounting to the $30\%$ of the data. 

Since data are simulated, both $\mathcal{X}_{b}$ and $\mathcal{X}_{bs}$ are, in fact, labeled. However, to mimic their use in a realistic setting where $\mathcal{X}_{bs}$ would represent the experimental, unlabeled, data, labels of $\mathcal{X}_{bs}$ have been employed for evaluating the quality of the results only. 

\begin{figure}[tb]
\begin{center}
\includegraphics[width=.37\textwidth, height=4.5cm]{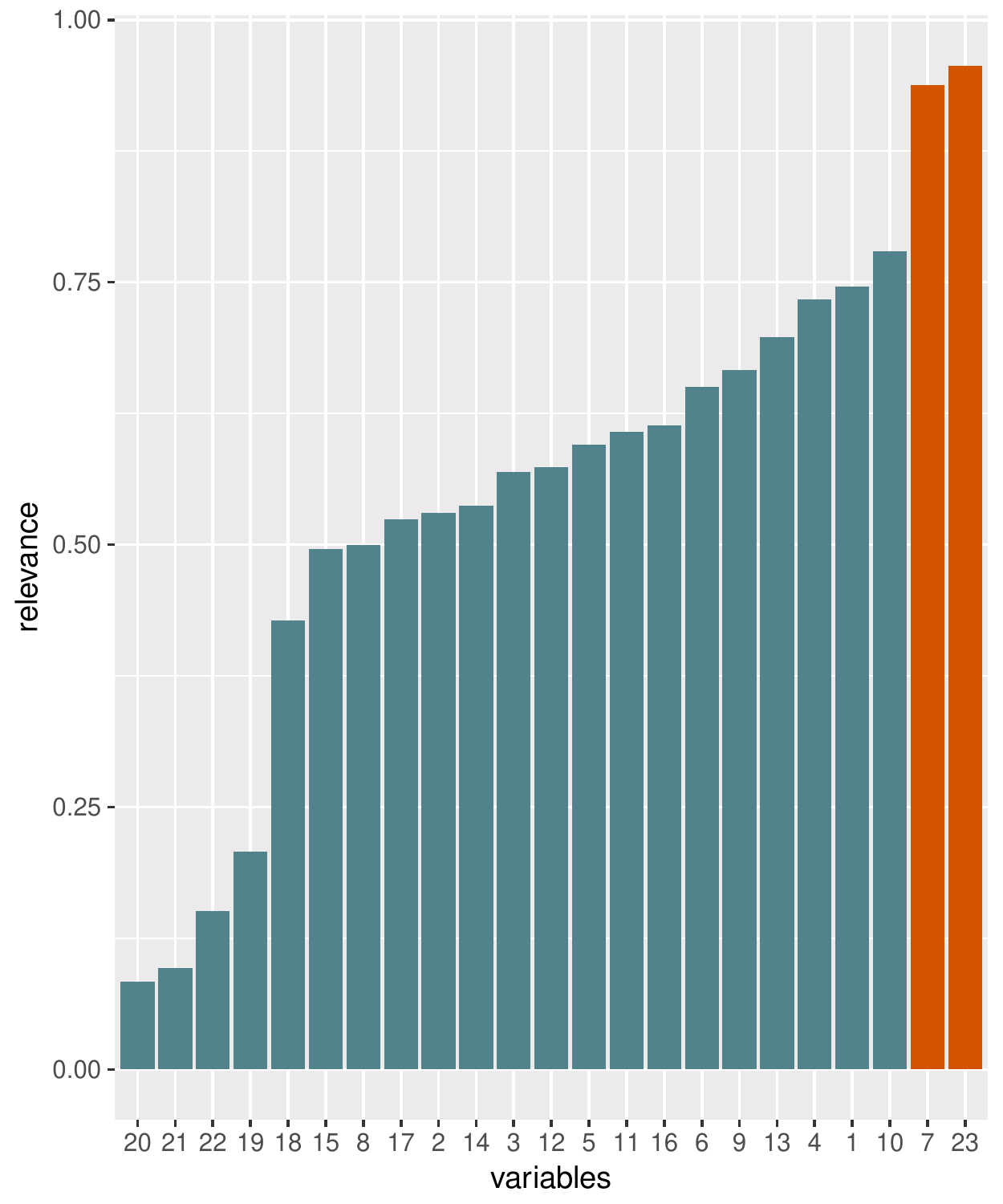}
\includegraphics[width=.32\textwidth,height=4.5cm]{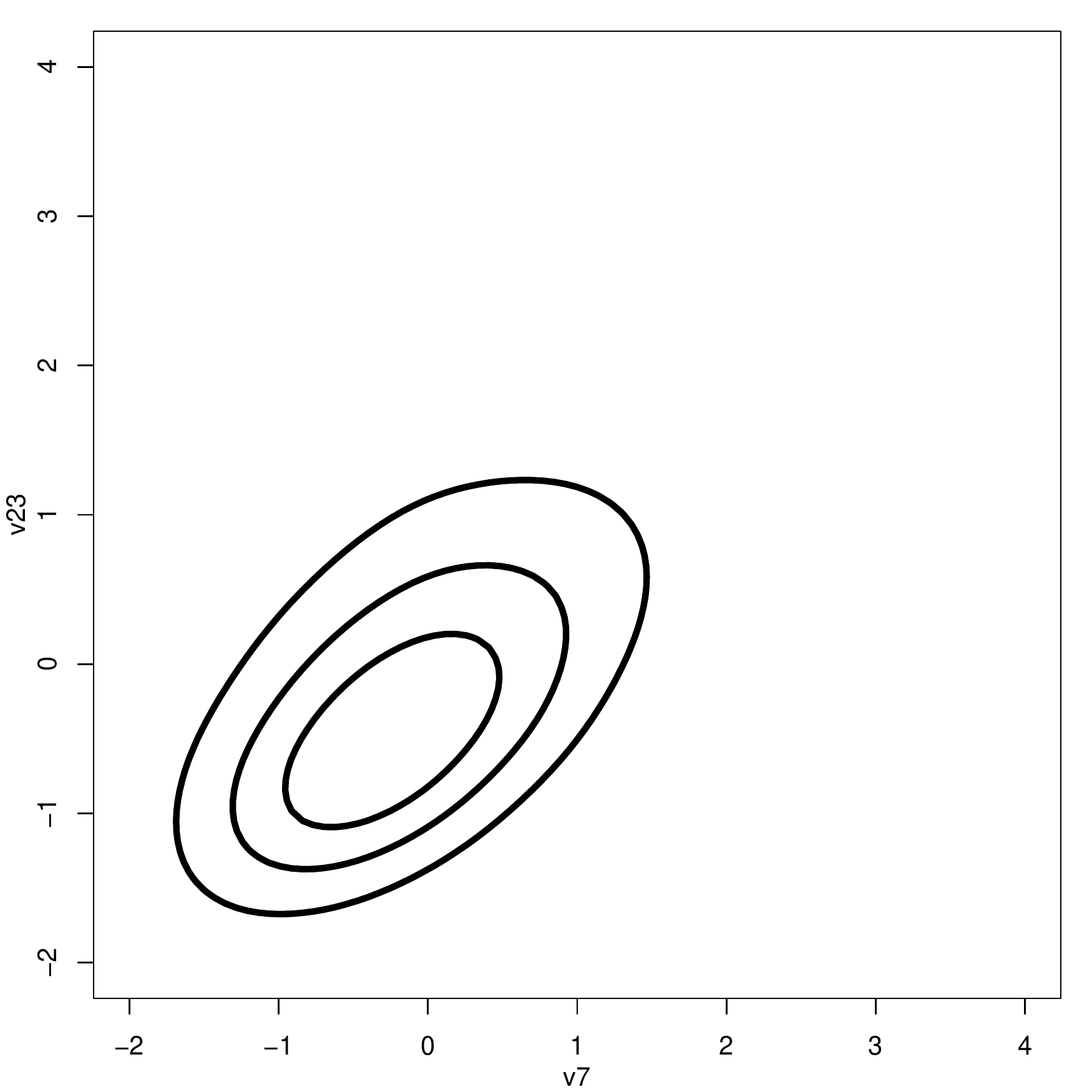}
\end{center}
\caption{Left: barplot of the relative informativeness of the whole set of considered variables, as resulting by the application of algorithm \ref{alg:varsel}. The orange bars on the right indicate the most relevant variables, selected for the subsequent steps of the procedure. %Middle: bootstrap stability distribution of the density estimate of the background versus the bandwidth value to estimate the density. Stability is measured by the integrated square difference (ISD) between the estimates built on different bootstrap samples from the background. The most stable not degenerate behaviour is associated to the kink of the plot, and it is coloured in orange. 
Right: contour plot of the background density estimate of the two selected variables.}\label{fig:varimp}
\end{figure}

In the left panel of Figure \ref{fig:varimp}, the results of the variable selection procedure are displayed. The test based on (\ref{eq:test}) has been performed by extracting $1000$ subsets of $k=3$ variables from the original 23. To compute the test statistic, we adopted the rule of thumb of selecting $h$ as asymptotically optimal for a Normal underlying background density. Two features, numbered as 23 and 7, and corresponding to the combined mass of two bottom quarks with two W bosons, and the transverse momentum of the leading jet show a remarkably different behavior between the background and whole process densities. In the subsequent analyses we have worked with these two variables only. The estimated background density on the reduced set of bivariate data $\mathcal{S}_b$ results unimodal, as illustrated in the right panel of Figure \ref{fig:varimp}. The estimate $\hat f_b$ has been obtained by selecting a plug-in gradient bandwidth $h_b$ \citep{Chacon_Duong2010}. It is worth noting, however, that due to the large amount of available data, most of automatic bandwidth selectors lead to grossly the same result. Due to the unimodality of $\hat f_b(\cdot; \mathcal{S}_b, \tilde h_b)$, the induced partition $\mathcal{P}_b(\mathcal S_b)$ is formed by one group only.

\begin{figure}[t]
\includegraphics[width=.3\textwidth]{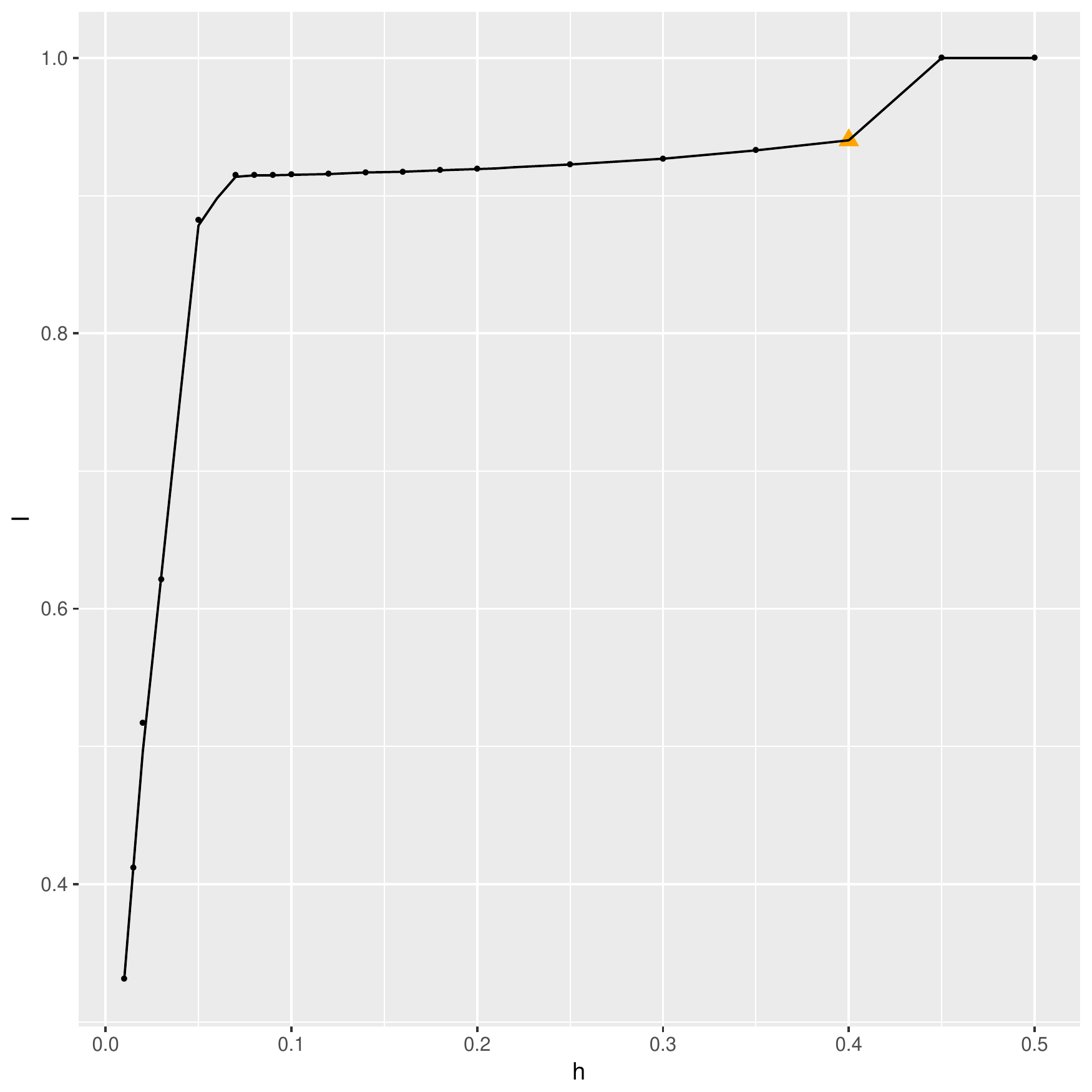}
\includegraphics[width=.3\textwidth]{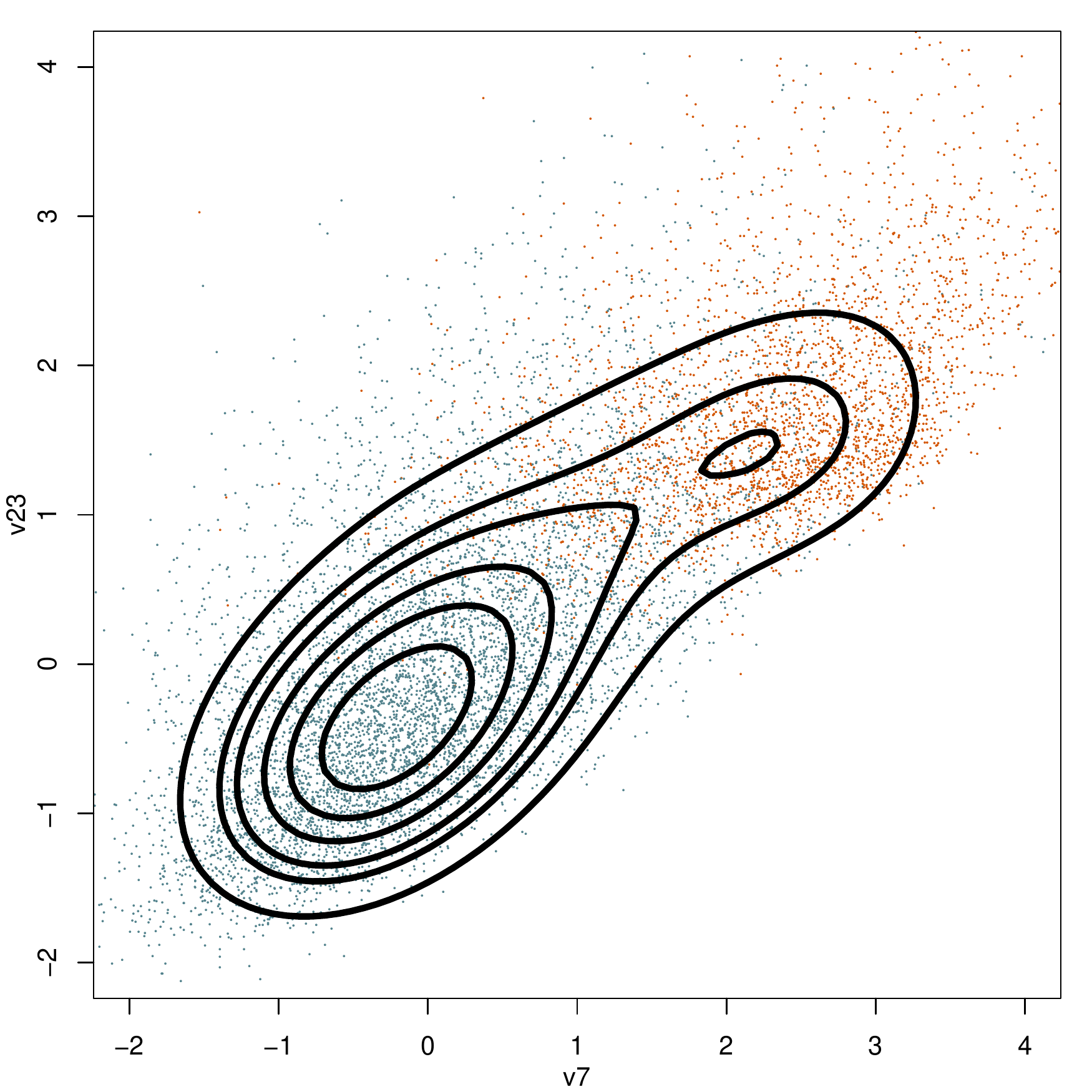}
\includegraphics[width=.3\textwidth]{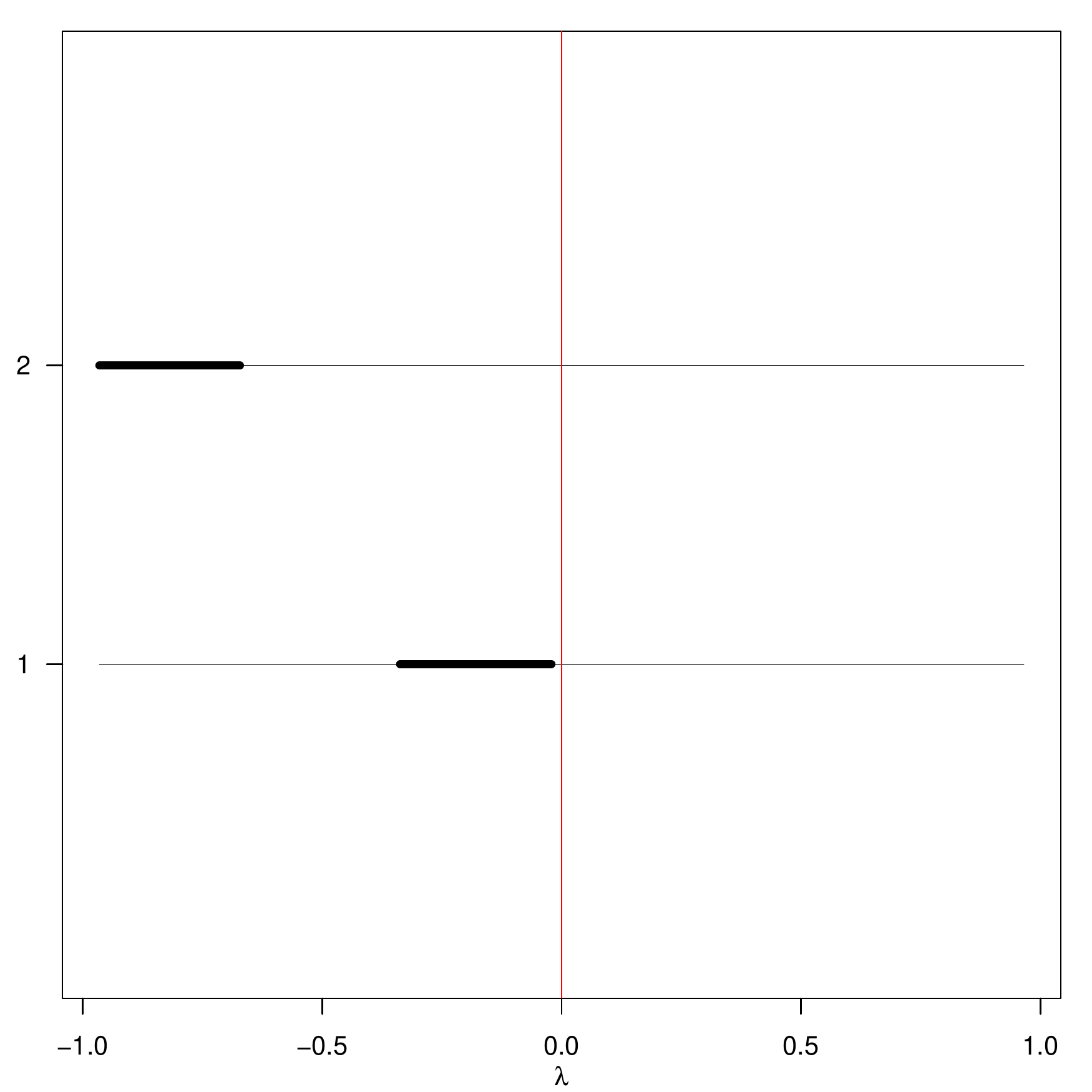}
\caption{Left: agreement index values of the classification between $\mathcal{P}_b(\mathcal{S}_b)$ and $\mathcal{P}_{bs}(\mathcal{S}_b)$ for varying badwidth selected to estimate $\hat{f}_{bs}(\cdot)$. The selected value $\tilde h_{bs}$ of $h_{bs}$ is highlighted with a triangular symbol in the plot and corresponds to the value leading to the maximum value of the agreement index when the number of modes of $\hat{f}_{bs}$ is larger than the number of modes of $\hat f_b$; Middle: contour plot of the density function estimated by selecting $\tilde h_{bs}$; Left: confidence intervals at the level $1- \alpha= 0.0001$ for the eigenvalues of the Hessian at the signal mode.}\label{fig:hbs}
\end{figure}

Figure \ref{fig:hbs} shows the results of the application of the procedure sketched in the Pseudo-algorithm \ref{alg:selhbs}. Nonparametric clustering has been performed by applying the mean-shift algorithm which allows for a natural classification of the background observations not employed to determine the partition. As an agreement measure, we have considered the Fowlkes-Mallows index, as it is sensitive to a different quality of partitions also when one of the two partitions is formed by one group only (as it is $\mathcal{P}_b(\mathcal S_b)$ in our case). The bandwidth $\tilde h_{bs}$ has been selected as the maximum value of the agreement index associated to more than one mode and determines a bimodal $\hat{f}_{bs}(\cdot, \mathcal{S}_{bs}, \tilde h_{bs})$ (middle panel of Figure \ref{fig:hbs}).
Specifically, the plot of the agreement index versus the bandwidth, illustrated in the left panel of Figure \ref{fig:hbs}, reveals that small values of the bandwidth $h_{bs}$ determine a low quality in the classification of the background observations, likely due to a heavily undersmoothed $\hat{f}_{bs}$ with a complex modality not shared by the background density. From some $h_{bs}$ on, the agreement index lifts up to high values, remaining stable for a wide range of $h_{bs}$, and associated to a bimodal $\hat f_{bs}$. This is a first rough indication about the non-spuriousness of the detected modes. Finally, $I$ grows up to its maximum value, occurring when the density estimate of the whole process gets unimodal as the background is. The significance of the additional mode, potential candidate to be the signal, is formally evaluated via the application of the test proposed by \citet{peronepacifico}. Inference relies on computing a bootstrap-based confidence interval for the eigenvalues of the Hessian at the given mode of the density estimate, based on a test sample of data (the selected $\mathcal{X}_{bs}^{T}$ in our case). The derived intervals, at the level $1-\alpha = 0.0001$ are entirely included in the negative semi-axis, suggesting the significance of the mode (right panel of Figure \ref{fig:hbs}).

Hence, density $\hat{f}_{bs}(\cdot, \mathcal{S}_{bs}, \tilde h_{bs})$ has been employed to determine a partition of $\mathcal{S}_{bs}$ in order to finally identify the signal events. Table \ref{tab:tables} compares the detected labels with the true ones and shows a satisfying quality of the partition, with a Fowlkes-Mallows index equal to 0.84 and a True positive rate amounting to the 80\% of the observed signal. %For the sake of evaluation, Table \ref{tab:tables} also includes the results of the classification of the test observations, not used to select $\tilde h_{bs}$

As a benchmark procedure, we also applied the parametric semisupervised method proposed by \citet{vatanen2012}. Due to the high dimensionality, data have been preliminarly reduced according to two different routes: first, we followed the authors suggestions and performed principal component analysis. We kept 6 components, to exceed the 50\% of explained variance. While in Table \ref{tab:tables} we just reported the aggregated background and signal classes, the method finds 12 background clusters and 4 additional components capturing the signal. The overall Fowlkes-Mallows index is equal to $0.77\%$, which is pretty satisfying but the true positive rate amounts to the $50\%$ only.
As a second route to reduce the data dimensionality, we considered the two variables selected according to the proposed nonparametric procedure. 
Results, reported in Table \ref{tab:tables}, are improved over the use of principal components analysis to reduce data dimensionality, thus showing further evidence about the relevance of the selected variables. However, the final partition is less accurate than the one obtained via the proposed nonparametic methodology. In this setting, 5 Gaussian components have been selected via the Bayesian Information Criterion to model the background and 4 components are used to fit the signal. 
  
All the analyses has been performed in the R environment \citep{R}, with the aid of the \verb,ks, package \citep{ks} to perform density estimation and nonparametric clustering based on the mean-shift algorithm.

\begin{table}[tb]
\caption{True classification of background and signal versus: classification obtained with the proposed nonparametric semisupervised procedure applied on the two selected variables, classification obtained with the parametric procedure proposed by \citet{vatanen2012} applied on the first 6 principal components and classification obtained with the same parametric method applied on the same two variables selected by the nonparametric approach.}\label{tab:tables}
\begin{center}
\begin{tabular}{p{2cm}}
\hline
\\
\hline
background \\
signal\\
\hline
\end{tabular}
\begin{tabular}{rrr}
 \multicolumn{3}{p{3cm}}{\small Nonparametric method}\\
  \hline
 & 1 & 2 \\ 
  \hline
  & 6582 & 441 \\ 
  & 604 & 2373 \\ 
   \hline
       \multicolumn{2}{l}{FMI}&0.84\\
          \multicolumn{2}{l}{TPR}&0.80\\
\end{tabular}
\begin{tabular}{rrr}
 \multicolumn{3}{p{2.5cm}}{\small Parametric method - 6 PC}\\
  \hline
 & 1 & 2 \\ 
  \hline
 & 6709 & 314 \\ 
   & 1500 & 1477 \\ 
   \hline
       \multicolumn{2}{l}{FMI}&0.77\\
       \multicolumn{2}{l}{TPR}&0.50\\
\end{tabular}
\begin{tabular}{rrr}
 \multicolumn{3}{p{2.5cm}}{\small Parametric method - 2 var}\\
  \hline
 & 1 & 2 \\ 
  \hline
 & 6646 & 377 \\ 
   & 1305 & 1672 \\ 
   \hline
       \multicolumn{2}{l}{FMI}&0.78\\
       \multicolumn{2}{l}{TPR}&0.56\\
\end{tabular}
\end{center}
\end{table}

\section{Final remarks} \label{sec:discussion}
%In this paper we have presented a global semisupervised nonparametric methodology aiming to detect a possible signal, behaving as a deviation from a known background process. In particular the proposed method addresses the issue of identifying anomalies when they appear collectively in regions of the sample space being compatible with the domain of the background.
%A QUESTO PUNTO, FORSE TOGLIEREI DEL TUTTO LA SECTION, visto che i concetti principali e le "giustificazioni" sono già state date. E terrei solo due commenti finali direttamente nella section precedente. 

In this paper we have presented a global semisupervised methodology aiming at identifying a possible presence of a signal, within the distribution of a known background process. While finding motivation in particle physics applications, the methodology easily extends to all those other fields where the searched signal represents any anomaly expected to appear collectively in regions of the sample space which are compatible with the domain of the normal process. 

Whatever application field is considered, an implicit common denominator is the greatest interest in the unknown signal process, whose detection would represent a far-reaching discovery. While any such discovery 
cannot be claimed without further analysis, the proposed methodology shall be considered as a fundamental step in the direction of forewarning of the possible presence of some anomaly, with the additional indication of which specific observations are the suspected anomalous ones. 
% be seen as an exploratory tool useful to gain some insights on the possible presence of a signal in the data.  In this way, guided by physical knowledge, it is possible to examine further the region of the sample space where the mode is located in order to gain a deeper understanding of the behavior. 
In this perspective, the proposed methodology has proved remarkably useful and its application to physical data has led to promising results, which overperform the parametric counterpart \citep{vatanen2012}. In addition to the anomaly detection, the proposed procedure for variable selection exhibits good results as well, with respect to standard alternatives such as principal components, in building a meaningful subspace to work on. % and the overall results overperform the main competitor. %In our opinion, considering a nonparametric approach, we introduce a great flexibility allowing to describe arbitrary shapes of the background process and of the unknown signal.  

\section*{Acknowledgements}
This project has received funding from the European Union's Horizon 2020 research and innovation programme under the Marie Sk\l odowska-Curie grant agreement No 675440 AMVA4NewPhysics. A preliminary version of this work appears in the proceedings: Casa, A., Menardi, G. (2017) Signal detection in high energy physics via a semisupervised nonparametric approach. Proceedings of the Conference of the Italian Statistical Society ``Statistics and data Sciences: new challenges, new generations''. Firenze. ISBN: 978-88-6453-521-0.

The authors wish to thank T. Dorigo (INFN) for the useful discussions and G. Kotkowski for the R code to implement the parametric method proposed by \citet{vatanen2012} and run for comparison.

\iffalse
{}
\fi

\bibliographystyle{plainnat}
\bibliography{biblio} 

\end{document}